\documentclass{sf2a-conf2024}
\usepackage{graphicx}
\usepackage{hyperref}
\usepackage{wrapfig}
\usepackage[]{natbib}  
\usepackage{epstopdf}
\usepackage{amsmath}

\def\BibTeX{{\rm B\kern-.05em{\sc i\kern-.025em b}\kern-.08em
    T\kern-.1667em\lower.7ex\hbox{E}\kern-.125emX}}
\bibpunct{(}{)}{;}{a}{}{,}  

\begin{document}

\TitreGlobal{SF2A 2024}


\title{SF2A Environmental Transition Commission:\\ Towards a desirable future for research in astronomy}

\runningtitle{Review SF2A-2024 S00}

\author{F.~Cantalloube}\address{Univ. Grenoble Alpes, CNRS, IPAG, F-38000 Grenoble, France}
\author{C.~Noûs}\address{Laboratoire Cogitamus, \url{https://www.cogitamus.fr/}}

\setcounter{page}{237}


\maketitle


\begin{abstract}
During its annual conference in 2024, the French Society of Astronomy \& Astrophysics (SF2A) hosted a special session dedicated to discussing the \emph{environmental transition} within the scope of our occupation. Since 2021, thinking on this subject has progressed significantly, both quantitatively and qualitatively. This year was an opportunity to take stock of the main areas of reflection that we need to keep in mind in order to implement a fair, collective and effective environmental transition. This proceeding summarizes the key points from the plenary session related to the  \emph{environmental transition} special session. 
The purpose of the messages disseminated here is to suggest ideas for reflection and inspiration, so as to initiate, stimulate, and foster discussions within the A\&A research community, towards the implementation of concrete measures to mitigate our environmental footprint. 
\end{abstract}



\section*{Foreword}
\textit{In 2024, the SF2A conference was held in Marseille, the second largest French city, located on the Mediterranean Sea coast. This preface gives an overview of the current environmental situation in this region, where discussions about the long-term prospects of our field, research in astronomy and astrophysics, have been raised.} 

The Mediterranean basin encompasses 23 countries on 3 continents, gathering a particularly dense population of 150~million people, showing wide disparities in access to resources, wealth and political situations. The region also receives 200~million tourists every year, which is one third of the world tourism. 
By its geographical situation, this region is a place of rich cultural and commercial exchanges, a major driver of demographic and economic growth. 
But the excessive exploitation of its abundant natural resources, makes the Mediterranean basin experiencing significant damages and irreversible losses affecting both terrestrial and marine ecosystems.

The Mediterranean sea is almost fully enclosed and located at mid-latitude, with low tides and high evaporation (38\% saltier and 56\% hotter than oceans) balanced with Atlantic ocean waters crossing the Strait of Gibraltar. 
This humidity is transported by eastwards jetstream making the zone a particularly high-pressure atmospheric zone (therefore hotter and drier, with events of heavy rainfalls). 
The Mediterranean sea represents 0.66\% of the global ocean surface but hosts about 9\% of the marine biodiversity. 
The Mediterranean land represents 1.6\% of the global land surface but hosts about 10\% of land biodiversity. 
With this, the Mediterranean basin is the second biodiversity hotspot on the planet, which mainly relies on two endangered ecosystems covering 7\% of the submarine surface: posidonia meadows and corallig\`enes. 
With its specific geographic and geopolitical situation, the Mediterranean basin suffers from the consequences of climate change, biodiversity loss and pollution at a higher rate than most regions worldwide \cite[MedEC,][]{MedECC}. 

The sea surface warming is estimated at $0.7^{\circ}$C per decade (a warming 20\% higher than oceans) and its acidity is rising at 7\% per decade. In August 2024, the median sea surface temperature (SST) reached a record of $28.9^{\circ}$C, rising the SST anomaly by almost $6^{\circ}$C. 
On top of the SST rising, increase of salinity, decrease of oxygen concentration, increase of $\text{CO}_2$ concentration, acidification, sea level rising, sedimentation and intensification of mineral dust coming from Sahara, are all direct and indirect consequences of climate change, putting extreme pressure on the biodiversity. 
Human activities such as aquaculture, canals and dams are also breaking migrating barriers, putting the Mediterranean basin ecosystem under high pressure (several tens of invasive species are on the black list of the International Union for the Conservation of Nature, \href{https://iucn.org/}{IUCN}), already fragilized by over-fishing (about 90\% of fish stocks are fished above sustainable levels). 
In addition, plastic pollution of the Mediterranean waters by micro-plastic\footnote{Fragments smaller than 5mm.} ($\sim130,000$ tons/year) and macro-plastic ($\sim500,000$ tons/year) is exacerbated by poor waste management due to tourist pressure and lack of strong political actions, added to its almost closed configuration. Other types of pollution due to human activities have disastrous effects: oil spills (400,000 tons/year), chemicals (such as sunscreen, medication components, ship bottom paints and other industrial products, all including endocrine disruptors), submerged ammunition, heavy metals, pesticides, herbicides and fertilizers used in agriculture such as Nitrate and Phosporous (notably provoking eutrophication) running-off into the sea during heavy rainfalls typical of Mediterranean climate. 
Noise pollution caused by shipping and other human enterprises can travel long distances, and marine species who may rely on sound for their orientation, communication, and feeding, can be harmed. 
Altogether, this triggers a steep loss of biodiversity and abundance (mass mortality of corals, changes in phytoplankton and zooplankton production disrupting food chains, competition with invasive species, and proliferation of species such as jellyfish), while making the ecosystem less resistant and resilient to increasing epizooties and zoonose.

The current land surface temperature in the Mediterranean region is $1.5^{\circ}$C above the pre-industrial level (a regional warming about 20\% above the global average). Climate change also leads to less rainfall (up to 30\% reduction by 2100), continued sea level rise ($0.5$ to $2.5~\mathrm{m}$ by 2100, leading to substantially drier and hotter conditions), coastal flooding, perturbation of the regional hydrological cycle, and rise in  extreme events in the region (forest fires, storms, sudden heavy rainfalls, etc.). The population density growth concurred to the destruction of natural environments with the artificialization of soil, fragmentation of habitat, and soil erosion. The littoral is also significantly affected by atmospheric pollution (essentially aerosols due to pollution generated by diesel engines from cruise and container ships), but also noise and light pollution. 

\textit{It is in such disastrous context, that in June 2024 the SF2A annual conference took place to discuss the ins and outs of French research in astronomy and astrophysics. How can we steer our professional activities in light of the current circumstances? How to think through theses issues and guide our future decisions?}


\section{Background and Framework}
From its inception, astronomy has been a holistic science, historically providing references to space, time, and spirituality. 
As such, astronomy has contributed greatly to the development of human societies (mainly through population migration, agriculture/settlement, and the telling of cosmogonies feeding any form of art). 
Modern astrophysics (beginning more or less with the birth of spectroscopy in the $\mathrm{18^{th}}$ century) is a multi-scale and multi-messenger science, originally based on observations of electromagnetic waves emitted from physical processes occurring above the Earth's atmosphere\footnote{By convention, the Earth atmosphere upper limit is defined at about 100~km altitude above sea level.}. 
Contemporary research in astrophysics involves a variety of methodologies, primarily: modeling, experiments (whether conducted in laboratories on Earth, in micro-gravity conditions in Earth orbits, or in situ for planetary science), and simulations. In this framework, instrumentation is a transverse activity linking these methodologies that lead to the interpretation of astronomical observations. This interpretation consists in arriving at a description of nature, playing with the cursor between simplification and completeness of the various parameters and processes at stake. 

Nowadays, the core of a researcher's work in astronomy and astrophysics is essentially based on data acquisition, analysis, and interpretation (concerning all the previously exposed methodologies). Other tasks are (a) teaching, training, supervision and mentoring, (b) communication within and outside the community via scientific publications, participation in conferences, public talks etc. and (c) community work such as organizing the community, reviewing papers or projects. Obviously, all these activities depend on resources, not only financial, but also in terms of allocated time, means and energy. 

In this context, the question is how to modulate the research activity in A\&A, facing the current environmental disaster and its numerous consequences on all facets of human societies. 
Many collectives, individuals and instances have been actively working on this question and the ones that arise from it. 
Below, we summarize some key questions, areas for reflection, and messages that, as a community, we may want to think about and address.

\section{The paradoxes}
In the Western societies, there is a strict correlation between the advancement of knowledge and the destruction of the Earth's ecosystem \citep[see e.g.][]{Dupont2024}. 
Furthermore, the study of astronomy and astrophysics is intertwined with a contextual social framework, as its progress is contingent on the advancements and acceptance by the society in which it is cultivated. 
As an illustration of this, there is a bidirectional exchange of technologies between commercial trade, military field, and fundamental sciences. 
Western societies, which have been at the forefront of progresses in understanding of astronomy and astrophysics, are still based on the ideology of human extrication from the Earth biosphere as a post-Cartesian way of thinking. As a whole, research in A\&A is not independent of the society in which it prospers, but is well and truly embedded in the social matrix. 
In what follows, we present some of the fundamental paradoxes that underlie contemporary research in A\&A.

\begin{itemize}
 \item  Carbon is the basic element of all known forms of life as we know it. It is also present on Earth in the form of inorganic (such as $\text{CO}_2$) and organic compounds (such as oil and natural gas). Without $\text{CO}_2$ gas in the atmosphere, the mean surface temperature of Earth would be too cold ($-18^{\circ}$C) for human societies to thrive in some regions. But the same $\text{CO}_2$, extracted from organic compounds and added to the troposphere by human industrial activities, is making the temperature of Earth surface warmer way too quickly, provoking mass mortality of many life forms on which the prosperity of human societies nevertheless depend.
  \item Now, during the first quarter of the $\mathrm{21^{st}}$ century, we reached a moment where astrophysicists are talking seriously about characterizing \emph{habitable exoplanets} and discovering life on \emph{habitable worlds}. Meanwhile, the habitability of Earth itself is threatened by human activities.  
  Scientists in other fields of the natural and social sciences are also struggling to get the societies responsible for this massive and abrupt destruction of the Earth's ecosystems to recognize and act against it. 
  \item To search for traces of life on extrasolar planets, astronomers have build telescopes installed in the most isolated and deserted areas, where conditions are so extreme that life has low chances of surviving, from the most arid desert of the Chilean Andes to space itself. 
  In addition, the locations chosen to host professional telescopes are often colonized territories \citep[e.g. the Hawaiian Islands ][the Canary Islands]{marichalar2021}, or territories that are not inhabited by the major countries operating the telescope (e.g. the Chilean Andes), or places with fragile habitats home to unique biodiversity (e.g. the Bure plateau in France). 
  \item With these telescopes, astronomers unveil stunning images of our universe, accessible in books, on the world wide web and displayed posters. However, as our industrial society and urban lifestyle expand, atmospheric and light pollution prevent access to the sensible experience of the sky. Stargazing is however a link with our terrestrial roots that gives us a sense of unity and profound link to the planet and what it harbours \citep[such as the feeling of wonder about the universe,][]{barragan2024}. 
  In addition, today's skies are spoiled by numerous aircraft and artificial satellites, especially the recently launched megaconstellations, adding a top-down light pollution that alters our feelings about the night sky. These surrounding objects, which cover the full Earth's sky, are too often contrary to any kind of consensual or democratic decision, and their commissioning is motivated solely by economic thirst. 
  \item The use of such ground- or space-based telescopes and probes brought mapping of the Moon, of Mars and other bodies in the solar system with unprecedented details. Meanwhile, only 10\% of the oceans below $200~\mathrm{m}$ have been explored. Therefore we do not know precisely the effect of various pollutions, biodiversity erosion and climate change on the oceanic ecosystems and its subtle equilibrium with adjacent biomes. 
\end{itemize}

Facing systemic issues, we recall that the consensual main goals of \href{https://en.wikipedia.org/wiki/Research}{research} are to serve the society as a whole, to bring progress to humanity (however ``progress" is defined) and to share the knowledge gained. Scientific research contributes to a better understanding of the world and inspires curiosity and wonder. This is for instance the stated introduction of the French Research Council (CNRS) front webpage: \emph{La recherche doit servir. Servir la société. Faire progresser l’humanité. Partager les savoirs. Innover. Et ainsi dépasser les frontières de la connaissance. Pour cela, le CNRS fait le pari d’une recherche mobilisant tous les domaines, en quête de progrès durable, au service d’une avancée technique, scientifique ou sociétale.}\footnote{https://www.cnrs.fr/fr/nos-recherches}

\section{Thinking keys}
The last few years have been filled with discussions about what research do we want for the field of Astronomy \& Astrophysics in a world drained by the environmental crises and its numerous consequences on human societies. The plenary contribution of the \emph{Commission Transition Environnementale} during the SF2A-2024 conference, was an opportunity to \href{https://journees.sf2a.eu/wp-content/uploads/2024/06/Cantalloube_S00.pdf}{summarize the key-points} of these discussions that are outlined below.

\paragraph{The meaning of progress} One first question that arises is what does \emph{progress} really mean, what do we call an \emph{innovation}? 
Research is constantly evolving field based on what is already known. Yet, there are so many examples of fundamental discoveries or technological breakthroughs that are either misused, or used too quickly before their overall deleterious effect on society is understood, proven and acted upon. Another point is the rebound effect, which shows that when a resource is optimized, it tends to be used more, so that there is no total gain. A technological boom is too often associated with monetary gains and profits, to which our Western society is devoted. This often breaks any attempt to take the time to evaluate, as a whole, how these findings get used. Even in astronomy this question remains relevant with the development of detectors, spectrometers, adaptive optics or image processing. And even from a very general point of view, what are we communicating to society with astronomy-related findings such as the notion of exoplanets in the so-called \emph{habitable zone}?

\paragraph{The duty to set an example} 
When talking about exemplariness of research and researchers, which legitimacy do astronomers have?  Astronomers consume a lot of resources by using infrastructures such as telescopes, providing a huge amount of data to store and analyse and by running massive simulations. What credibility do we have in an atomized and increasingly untrusting society?
One of the key points is to share knowledge more effectively between different, often kept apart, fields of research (such as the environmental sciences, but even more so with social sciences) for the benefit of humanity. In particular for astronomers, it means forging closer links with the social sciences. 
And more generally, perhaps we should strive to strengthen links with journalists, politicians, artists and society as a whole so as to co-construct an effective fight against misinformation and lobby violence, while remaining humble in the face of our state of knowledge. 

\paragraph{The increasing pace of research}
Research is following the extractivist model followed by our growth-oriented capitalist societies. Astronomers, like most researchers, aim for more: more data, more instruments, more papers, more PI-ship, more collaborations, more conferences... This pace is not in line with the lack of recruitment throughout the sector and is simply not sustainable: the number and the scale of project is slowly converging towards the physical capacity of our community and adjacent industrial sector. This notably has a deleterious effect on mental health with an increase of \emph{burnout} symptoms in academia. 

\paragraph{Individualism in research}
Research has never been a lone pursuit, contrary to popular discourses and told stories exacerbated by the epinal image of a scientist (also shown with the concept of Nobel prize usually awarded to individuals). Without a sufficient surrounding (access to basic needs such as food and healthcare, absence of conflicts etc.), no genuine discovery can be made. 
Our society rewards and claims meritocracy but more and more studies show that this simply does not exist. 
The current policy in academic promotes competition and individualism: getting grants, being a PI, leading or coordinating groups etc. It is however possible to get out this predatory model by supporting collective work towards redistributing resources in a fair way.

\paragraph{Decision power and democracy}
Most of the time, people claim their lack of power and choice in political decisions. On the level of academic decisions, we are also often poorly in power of the choices taken by the institutions upon which our activities depend. And even more so in a context of competition with other countries whose operating and funding practices may be very different. We however hold a substantial collective weight that we may use wisely to recover a sense of democracy, in our society in general, and in the research ecosystem in particular, using for instance concepts of \emph{collective intelligence}.

\paragraph{Links and divides}
If we are to move forward serenely and effectively, together as a community, we must remain vigilant to the fractures that punctuate our community and which can be a source of inequalities. Gender, generation, social class and/or cultural background can all have an impact on how we feel and perceive, and how do we act on changes in the way our society operates. 
On the contrary, many links and agreements already exist, based on a common culture in our profession, which brings us together. We may preserve and improve this connection in order to choose directions that suit as many people as possible.  

\paragraph{Assessing research}
A question that arises from many discussions is on how to assess the \emph{quality} of research at any career stage. In today's dominant society, we tend to quantify everything, some argue for the sake of simplicity and some argue that it is about control. In all cases, it is an extremely reductive view of the influence of a researcher's activities. Numerous examples show that the pressure to quantifying research results provoke a loss of creativity. Accept that prejudice exists, that researchers are also prone to biases, and fight against discrimination is perhaps healthier, and researchers could work together towards more qualitative evaluations. This requires to take more time to assess research, but on the long run, it may be beneficial for research.

\paragraph{The values of research}
The latter thoughts also touch upon the question of our society's values and their hierarchization. For example, research on living organisms has long been ignored or even despised in the utilitarian context of the 1950s, which saw the birth of the \emph{great acceleration}. We are barely aware of the inter-species and intra-species interactions or the effects of domestication in the massive decline and erosion of biodiversity that is underway. In the same way, women's bodies have long been rejected in medical research, the standard being the Caucasian man, and numerous errors and delays have been the consequences, despite the fact that these effects concern not a minority but simply half of all human beings. 
In addition to this bias, does it really matter if our research is useful? Is art (apart from design, that is) useful? In this direction, is astronomy more valuable than other sciences (because it's ancient, it touches upon the questions of the origins of life and so on)? To go further, what does justify the resources and policies implemented in the various sectors; what interests these different sectors do serve?

\paragraph{The obligation of neutrality} 
In our profession, any issue that questions the inequalities or discriminations that fragment our world is often brushed aside and avoided because it touches on the notion of \emph{neutrality} of researchers. However, the concept of neutrality is a well-known delusion for researchers: it cannot exist because every decision taken is guided by some conditioned beliefs. So we might as well face up to it and embrace it: re-colouring our ideas, thoughts and expressions with subjectivity. And anyway, in the face of the various crisis hitting our world, do we really want to be neutral? Do we really need to be neutral? Isn't it our duty not to be? 
The recent notice published in 2022 \citep{comets2022}, written by members of the CNRS Ethic Committee (COMETS), gives an open reflection on these aspects, with an emphasis on the environmental issues.

\paragraph{Ethic and integrity of research projects} 
This same line of thinking fuels the notion of \emph{research ethics}: why and how can we pursue our research activities while guaranteeing a minimum commitment to ethics and integrity. 
Our community is inflected by colonialism, paternalism, patriarchy, racism, classism, and so on. How can we remain attentive to these social issues in which our community is embroiled? 
How do we refuse or break away from prejudicial projects? Astronomy, with its ability to inspire the general public, is often used as a cover to whitewash certain environmentally damaging activities or military industries (sometimes dubbed as \emph{astrowashing}). What compromises do we have to make to go against certain harmful projects that nonetheless bring us funding and opportunities to advance our research and career? How can we be coherent about the fact that funding sometimes comes from private partners with diverging economic interests?

\paragraph{Revisiting the meanings of research}
When doing fundamental research, are we really exploring or are we fleeing our own reality? 
Is the study of natural sciences just a way to avoid the numerous social problems that we face? Are we giving meaning or are we lying to ourselves and hiding from our own powerlessness? 
We should ask ourselves these fundamental questions in the light of the major changes we are facing if we are to be clear about our roles, our means of action, and our sphere of influence.

\vspace{0.5cm}
Keeping all these key points in mind, how can we together build a desirable future for our community: accepting a massive paradigm shift, rethinking the notion of progress and values, and nurturing the hope of improving our society. 
While maintaining and improving the quality of our scientific research, preserving good working practices, protecting our unique academic freedom, and cultivating well-being within the research ecosystem? 
While promoting the resilience of this system through diversity, equity, inclusion and accessibility in our disciplines? 
While reshaping our roles within society, breaking the division with social sciences, art, policy, and communication media?

\section{Conclusions}
Many individuals, including artists and scientists, share the same concern: without a global, rapid and ambitious action to fight against climate change and environmental degradation, we are likely to lose most of the wonders of the planet Earth hosting our human societies. Everything we know and feel is enclosed within a thin layer of merely a few hundred kilometers around the surface of oceans surrounding the planet. 
In this context, it is of utter importance to perpetuate the enchantment and curiosity that drive us, inhabitants of Earth, to protect our world. 
Fundamental research, and especially astronomy, is often a source of awe and inspiration, which serves as a reminder that all known life forms are hosted on a single planet governed solely by the laws of physics. 
Research is one among many aspects that makes the society we built singular, thrilling to comprehend nature and its origins towards improving our humanity. 

Astronomers are witnesses to the problem, are part of the problem, and suffer from the problem. But astronomers also stand a chance to be part of the potential solutions. Before the constraints due to changing environmental conditions become too strong, there is now an opportunity to reshape and improve the ways of thinking, doing and using science. The transformation of our societies should be seen as an intellectual stimulus and therefore as an important component of our research efforts \citep{ripple2024}. In conclusion, it is now advisable to infuse all these questions and make them our own, to keep them in mind as we interact with people inside and outside our community, to incorporate them into our decision making, and finally, to make them an integral part of our profession and identity. 



\begin{figure}[!h] 
\centering
\includegraphics[scale=0.18]{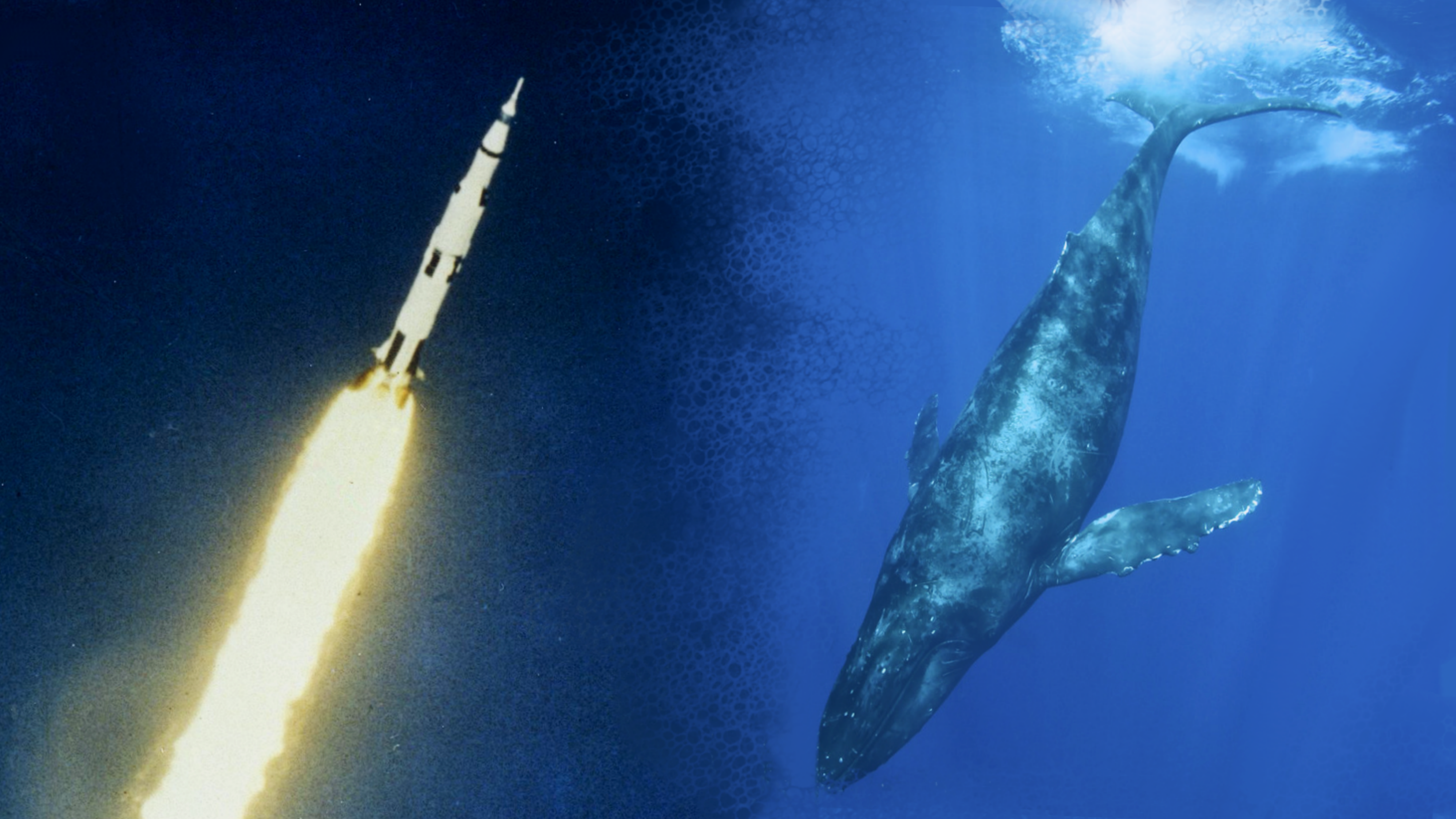}
\caption{Image shown during the presentation to illustrate the correlation between industrial developments and ecosystems destruction: the profitability of whale oil at the end of the $19^{th}$ century, used mainly as fuel, justified the slaughter of these large cetaceans and is one of the main causes of their decline. Without this massacre, technological growths such as those that allowed a handful of U.S. citizens to land on the moon and return safely, would presumably not have occurred.}
\end{figure}



\begin{acknowledgements}
We would like to thank the SF2A board members for giving us the opportunity to express ourselves freely during a plenary session. 
We also thank B.~Pope, J.~Milli, G.~Chaverot and F.~Malbet for reading through, offering suggestions for improvement, and supporting the angle of expression chosen in this proceeding.
\end{acknowledgements}

\bibliographystyle{aa}  
\bibliography{Cantalloube_S00} 

\end{document}